\documentclass[a4paper]{jpconf}
\usepackage{amsfonts,latexsym}
\usepackage{amsmath}
\usepackage{amscd}
\usepackage{float,amsmath,amssymb,mathrsfs,bm,multirow,graphics}
\usepackage[dvips]{graphicx}
\usepackage[percent]{overpic}
\usepackage{enumerate}

 \newcommand{\nequation}{\setcounter{equation}{0}}
\newcommand{\R}{{\Bbb R}}

\newcommand{\C}{{\Bbb C}}

\newcommand{\Z}{{\Bbb Z}}

\DeclareMathOperator{\im}{Im}
\DeclareMathOperator{\re}{Re}

\def\XXint#1#2#3{{\setbox0=\hbox{$#1{#2#3}{\int}$}
\vcenter{\hbox{$#2#3$}}\kern-.5\wd0}}

\allowdisplaybreaks

\newtheorem{theorem}{Theorem}[section]

\newtheorem{definition}[theorem]{Definition}

\newtheorem{example}[theorem]{Example}

\begin{document}

\title{Perturbative and Exact Results on the\\ Neumann Value for the Nonlinear
  Schr\"odinger\\ Equation on the Half-Line}
\author{A. S. Fokas}
\address{Department of Applied Mathematics and Theoretical Physics, University of Cambridge, Cambridge CB3 0WA, UK, and Research Center of Mathematics, Academy of Athens, 11527, Greece.}
\ead{T.Fokas@damtp.cam.ac.uk}

\author{J. Lenells}
\address{Department of Mathematics and Center for Astrophysics, Space Physics \& Engineering Research, Baylor University, One Bear Place \#97328, Waco, TX 76798, USA.}
\ead{Jonatan\_Lenells@baylor.edu}

\begin{abstract}
  The most challenging problem in the implementation of the so-called
  \textit{unified transform} to the analysis of the nonlinear Schr\"odinger
  equation on the half-line is the characterization of the unknown boundary
  value in terms of the given initial and boundary conditions. For the
  so-called \textit{linearizable} boundary conditions this problem can be
  solved explicitly. Furthermore, for non-linearizable boundary conditions
  which decay for large $t$, this problem can be largely bypassed in the sense
  that the unified transform yields useful asymptotic information for the large
  $t$ behavior of the solution. However, for the physically important case of
  periodic boundary conditions it is necessary to characterize the unknown
  boundary value. Here, we first present a perturbative scheme which can be
  used to compute explicitly the asymptotic form of the Neumann boundary value
  in terms of the given $\tau$-periodic Dirichlet datum to any given order in a
  perturbation expansion.  We then discuss briefly an extension of the
  pioneering results of Boutet de Monvel and co-authors which suggests that if
  the Dirichlet datum belongs to a large class of particular $\tau$-periodic
  functions, which includes $\{a \exp(i \omega t) \, | \, a>0, \, \omega \geq
  a^2\}$, then the large $t$ behavior of the Neumann value is given by a
  $\tau$-periodic function which can be computed explicitly.
\end{abstract}


\section{Introduction}\nequation
Let $q(x,t)$ satisfy the nonlinear Schr\"odinger (NLS) on the half-line with a given initial condition,
\begin{align} \label{1.1}
& i q_t+q_{xx}-2\lambda|q|^2q=0, && 0<x<\infty, \quad t>0, \quad \lambda=\pm1,
	\\ \label{1.2}
& q(x,0)=q_0(x), && 0<x<\infty,
\end{align}
where $q_0(x)$ has sufficient decay as $x \rightarrow \infty$.

In the above setting, the so-called unified transform \cite{F1997} can be used to analyze problems with either a linearizable boundary condition or with a non-linearizable boundary condition which decays
for large $t$. 

\subsection{Linearizable bounday conditions}
Suppose that equations \eqref{1.1} and \eqref{1.2} are supplemented with any of the following boundary conditions:
\begin{subequations}
\begin{align}
& q(0,t)=0,
	\\ 
& q_x(0,t)=0,
	\\ 
& q_x(0,t)- \chi q(0,t)=0, \qquad \text{$\chi$ real constant}. 
\end{align}
\end{subequations}
Then, the unified transform yields a solution with the same level of efficiency as the celebrated inverse scattering transform \cite{F1997, F2002, trilogy1}.

The above problems can also be solved by the usual inverse scattering transform after mapping them to certain equivalent problems on the line \cite{BFS}. However, the unified transform can also be 
implemented for linearizable boundary value problems for integrable nonlinear PDEs involving a third order  spatial derivative---such as the KdV equation---which \textit{cannot} be mapped to appropriate 
problems on the line.

\subsection{Boundary conditions which decay as $t \rightarrow \infty$}
Suppose that equations \eqref{1.1} and \eqref{1.2} are supplemented with the Dirichlet boundary condition
\begin{equation}
\label{1.4}
g_0(t)=q(0,t), \quad 0<t<\infty,
\end{equation}
where $g_0(t)$ vanishes as $t \rightarrow \infty$. In this case, the unified transform expresses the solution $q(x,t)$ in terms of the solution of a $2\times 2$ matrix Riemann-Hilbert (RH) problem which 
is uniquely defined in terms of the \textit{spectral functions} $\{a(k),b(k),A(k),B(k)\}$. The first two of these functions can be obtained in terms of the given initial datum $q_0(x)$, however
\{$A(k),B(k)$\} depend on both $g_0(t)$ and on $g_1(t)=q_x(0,t)$. Thus, in order to compute $q(x,t)$ it is necessary to first characterize the so-called Dirichlet to Neumann map, i.e. to express $g_1(t)$ in terms of
$g_0(t)$ and $q_0(x)$. However, since $g_0(t)$ vanishes for large $t$, using the fact that the above spectral functions are $t$ independent, it is still possible to compute the large $t$-asymptotics of the above 
RH problem, and hence the large $t$-asymptotics of $q(x,t)$ without characterizing fully the spectral functions $A(k)$ and $B(k)$.

\subsection{The Dirichlet to Neumann map}
Two different approaches for analyzing the Dirichlet to Neumann map were recently presented in \cite{trilogy1} and \cite{trilogy2}. The formulation in \cite{trilogy1} is based on the analysis of the eigenfunctions involved in the definition of \{$A(k), B(k)$\} (see also \cite{DMS2001, F2005}), whereas the formulation in \cite{trilogy2} is based on an extension of the Gelfand-Levitan-Marchenko approach first introduced in \cite{BFS2003}.
 
\subsection{$\tau$-periodic boundary condition}
Suppose that equations \eqref{1.1} and \eqref{1.2} are supplemented by equation \eqref{1.4}, where $g_0(t)$ is a periodic function with period $\tau > 0$.

A perturbative scheme for computing the large $t$ behavior of $g_1(t)$ is introduced in \cite{trilogy2}. In particular, it is shown in \cite{trilogy2} for the NLS, and in \cite{HF2013} for the mKdV, that if $g_0(t)= \alpha \sin t$, $\alpha \in \mathbb{R}$, then it is possible to obtain the function $g_1(t)$ for the NLS and the functions \{$g_1(t), q_{xx}(0,t)$\} for the mKdV as a perturbation expansion in the parameter $\alpha$, up to and including terms of $O(\alpha^3)$. Furthermore, these functions are $2\pi$-periodic, at least up to this order. Unfortunately, the perturbative approach of \cite{trilogy2} is quite cumbersome and it is practically impossible to go beyond terms of $O(\alpha^3)$.

In the particular case that
\begin{align}\label{1.5}
g_0(t)= \alpha e^{i\omega t}, \qquad \alpha >0, \quad \omega \in \mathbb{R},
\end{align}
pioneering results for the focusing NLS have been obtained in a series of papers of Boutet de Monvel and co-authors \cite{BIK2007, BIK2009, BKS2009, BKSZ2010} by
employing the unified transform. In particular, it is shown in \cite{BKS2009} that for the focusing NLS, $\lambda=-1$, there exists a solution
$q(x,t)$, such that
\begin{align}\label{1.6}
g_1(t) \sim c e^{i\omega t}, \qquad t \rightarrow \infty, \quad c \in \mathbb{C},
\end{align}
if and only if the triple of constants \{$\alpha,\omega,c$\} satisfy either 
\begin{align}\label{1.7}
 \Big(c=\pm \alpha \sqrt{\omega-\alpha^2} \ \ \mathrm{and} \ \ \omega \geq \alpha^2\Big) \quad \mathrm{or} \quad \Big(c=i \alpha \sqrt{|\omega|+2\alpha^2} \ \ \mathrm{and} \ \ \omega \leq -6\alpha^2\Big).
\end{align}

\subsection{Outline of the paper}
Here we first present a new perturbative scheme for computing the large $t$ asymptotic behavior of $g_1(t)$ in terms of the $\tau$-periodic 
function $g_0(t)$. As illustrations, we consider the example of the single exponential \eqref{1.6}, as well as
the example of
\begin{equation}
\label{1.8}
g_0(t)= \alpha  e^{i\omega t}+ \beta  e^{-i\omega t}, \quad \alpha,\beta \in \mathbb{C}, \ \ \omega>0.
\end{equation}
The main difference between this new perturbative approach and the one used in \cite{trilogy2} is that in the latter approach the analysis was first
carried out for all $t$ and then the limit $t \rightarrow \infty$ was computed, whereas in the new approach the analysis is carried
out directly in the limit $t \rightarrow \infty$. This simplifies the relevant algorithm enormously, as illustrated by the fact that for the 
above two examples we give formulae up to and including terms of order eight.

We also discuss briefly an extension of the formalism introduced in \cite{BIK2007, BIK2009, BKS2009, BKSZ2010}, which suggests that using the technique of finite gap integration,
it is possible to obtain a large class of $\tau$-periodic functions which have the crucial property that if they are assigned as the Dirichlet
data for the NLS on the half line, then the associated function $g_1(t)$ has the property that it asymptotes as $t \rightarrow \infty$ to a
$\tau$-periodic function and furthermore this function can be computed explicitly.

\section{A new perturbative approach} \label{pertsec}
In the context of asymptotically $t$-periodic data, there are situations where we do not necessarily need to know the Neumann value $g_1(t)$ for all $t > 0$, but it is sufficient to know its asymptotic form $g_1^B(t)$. It is therefore natural to ask the following question: Given the asymptotic form $g_0^B(t)$ of the Dirichlet datum, can we find the asymptotic form $g_1^B(t)$ of the Neumann value? Theorem \ref{pertth} below provides a constructive algorithm for computing the asymptotic form of the Neumann value from the asymptotic form of the Dirichlet data in a perturbative expansion. In other words, it provides an explicit construction of the Dirichlet to Neumann map for asymptotically $t$-periodic data in the limit of large $t$ and small data.

\subsection{Main result}
We first define what we mean by a perturbative solution.
 
\begin{definition}\upshape
A {\it perturbative solution of the NLS equation (\ref{1.1}) in the quarter plane} is a sequence of smooth functions $\{q_N(x,t)\}_1^\infty$  defined for $x>0$ and $t>0$ with the following properties:
\begin{enumerate}[(i)]
\item The formal power series 
\begin{align}\label{qformal}
q = \epsilon q_1 + \epsilon^2 q_2 + \epsilon^3 q_3 + \cdots
\end{align}
satisfies (\ref{1.1}) in the quarter plane $\{x> 0, t > 0\}$ to all orders in a perturbative expansion, that is, 
\begin{align*}
O(\epsilon): & \quad iq_{1t} + q_{1xx} = 0,
	\\
O(\epsilon^2): & \quad iq_{2t} + q_{2xx} = 0,
	\\
O(\epsilon^3): & \quad iq_{3t} + q_{3xx} - 2\lambda |q_1|^2 q_1 = 0,
	\\
O(\epsilon^4): & \quad iq_{4t} + q_{4xx} - 2\lambda (q_1^2 \bar{q}_2 + 2q_1 q_2 \bar{q}_1) = 0,
	\\
 \vdots &
	\\
O(\epsilon^N): & \quad iq_{Nt} + q_{Nxx} - 2\lambda \big\{|q|^2 q\big\}_N = 0,
\end{align*}
where $\{ \cdots\}_N$ denotes the coefficient of $\epsilon^N$ of the enclosed expression.
	
\item For each $N$, $q_N(x,t)$ and all its partial derivatives have continuous extensions to $\{x\geq 0, t \geq 0\}$.

\item For each $N$, $q_N(\cdot,0)$ belongs to the Schwartz space $\mathcal{S}(\R_+)$.
 \end{enumerate}
\end{definition}

We can now state our main result.

\begin{theorem}\label{pertth}
Let $\{g_{0N}^B(t), g_{1N}^B(t)\}_1^\infty$ be periodic functions of period $\tau = \frac{2\pi}{\omega} > 0$.
Suppose $\{q_N(x,t)\}_1^\infty$ is a perturbative solution of NLS in the quarter plane such that, for each $N$, the Dirichlet and Neumann boundary values of $q_N$ asymptote towards $g_{0N}^B(t)$Ê and $g_{1N}^B(t)$ respectively in the sense that
\begin{align}\nonumber
& q_N(0, \cdot) - g_{0N}^B \in \mathcal{S}(\R_+), && N \geq 1, 
	\\ \label{qNgN}
& q_{Nx}(0,t) - g_{1N}^B(t) = O(t^{-3/2}), && t\to \infty,Ê\quad N \geq 1.
\end{align}
Suppose the set $\{q_N(\cdot, t)| t \geq 0\}$ is bounded in $L^1(\R_+)$ for each $N \geq 1$.

Then the asymptotic Neumann values $\{g_{1N}^B(t)\}_{N = 1}^\infty$ can be constructed explicitly from the asymptotic Dirichlet values $\{g_{0N}^B(t)\}_{N = 1}^\infty$ as follows. 
Let $\{a_{N,n}\}_{n=-\infty}^\infty$ denote the Fourier coefficients of $g_{0N}^B$:
\begin{align}\label{g0NBFourier}
g_{0N}^B(t) = \sum_{n=-\infty}^\infty a_{N,n} e^{in\omega t}, \qquad N \geq 1.
\end{align}
Then the Fourier coefficients  $\{c_{N,n}\}_{n=-\infty}^\infty$ of
\begin{align}\label{g1NBFourier}
g_{1N}^B(t) = \sum_{n=-\infty}^\infty c_{N,n} e^{in\omega t}, \qquad N \geq 1,
\end{align}
are given by
\begin{align}\nonumber
c_{N,n} = &\; \bigg\{ \bigg(
-2i\lambda k \sum_{l,m=-\infty}^\infty \bar{a}_l d_m(k) d_{n+l-m}(k) 
 - \lambda  \sum_{l, m=-\infty}^\infty  \bar{c}_{l} d_m(k) d_{n+l-m}(k)
	\\ \label{cNndef}
&+ 2\lambda \sum_{l,m = -\infty}^\infty \bar{a}_l a_m d_{n+l-m}(k) + 2ik a_n\bigg)\bigg|_{k = k_1(n)} \biggr\}_N,
\end{align}
where 
\begin{itemize}
\item $a_n = \sum_{N=1}^\infty a_{N, n} \epsilon^N$, $c_n = \sum_{N=1}^\infty c_{N, n} \epsilon^N$, and 
$d_{n}(k) = \sum_{N=1}^\infty d_{N, n}(k) \epsilon^N$. 

\item The coefficients $d_{N,n}(k)$, $N \geq 1$, $n \in \Z$, satisfy
\begin{align}\nonumber
& d_{N,n}(k) = \frac{1}{4ik^2 + in\omega}\biggl\{
-2\lambda k \sum_{l,m=-\infty}^\infty \bar{a}_l d_m(k) d_{n+l-m}(k) 
	\\ \nonumber
& + \lambda i \sum_{l, m=-\infty}^\infty  \bar{c}_{l} d_m(k) d_{n+l-m}(k)
- 2i\lambda \sum_{l,m = -\infty}^\infty \bar{a}_l a_m d_{n+l-m}(k) + 2k a_n + i c_n\biggr\}_N, 
	\\ \label{dNndef}
& \hspace{7cm} n \in \Z, \quad N \geq 1, \quad k \in \partial D_1^0,
\end{align}
where $D_1^0 = \{\re k > 0\} \cap \{\im k > 0\}$ denotes the first quadrant of the complex $k$-plane.

\item $k_1(n)$ denotes the unique root of $4k^2 + n\omega = 0$ in $\partial D_1^0$, i.e.
$$k_1(n) = \begin{cases} \frac{i\sqrt{n\omega}}{2}, & n \geq 0, \\
\frac{\sqrt{-n\omega}}{2}, & n < 0.
\end{cases}$$
\end{itemize}
\end{theorem}

We refer to \cite{tperiodic} for a proof of Theorem \ref{pertth}. In the remainder of this section, we explain how the $c_{N,n}$'s are determined from equations (\ref{cNndef})-(\ref{dNndef}) and provide several examples. 

\subsection{Construction of the $c_{N,n}$'s}
Equation (\ref{cNndef}) with $N = 1$ yields
\begin{subequations}\label{c1nd1n}
\begin{align}\label{c1n}
  c_{1,n} = 2i k_1(n) a_{1,n}, \qquad n \in \Z.
\end{align}
Substituting this into equation (\ref{dNndef}) with $N = 1$, we find
\begin{align}\label{d1n}
d_{1,n}(k) = \frac{2k a_{1,n} + i c_{1,n}}{4ik^2 + in\omega} 
= \frac{a_{1,n}}{2i(k+ k_1(n))}, 
\qquad n \in \Z, \quad k \in \partial D_1^0.
\end{align}
\end{subequations}
Similarly, equations (\ref{cNndef}) and (\ref{dNndef}) with $N = 2$ yield
\begin{subequations}\label{c2nd2n}
\begin{align}\label{c2n}
& c_{2,n} = 2i k_1(n) a_{2,n}, && n \in \Z,
	\\ \label{d2n}
&d_{2,n}(k)  = \frac{a_{2,n}}{2i(k+ k_1(n))}, && n \in \Z, \quad k \in \partial D_1^0.
\end{align}
\end{subequations}
Continuing in this way, equations (\ref{cNndef}) and (\ref{dNndef}) with $N = 3$ yield
\begin{subequations}\label{c3nd3n}
\begin{align} \nonumber
c_{3,n} = &\; \bigg(
-2i\lambda k \sum_{l,m=-\infty}^\infty \bar{a}_{1,l} d_{1,m}(k) d_{1,n+l-m}(k) 
 - \lambda \sum_{l, m=-\infty}^\infty  \bar{c}_{1,l} d_{1,m}(k) d_{1,n+l-m}(k)
	\\ \label{c3n}
&+ 2\lambda \sum_{l,m = -\infty}^\infty \bar{a}_{1,l} a_{1,m} d_{1,n+l-m}(k) + 2i k a_{3,n} \bigg)\bigg|_{k = k_1(n)}, \qquad n \in \Z,
	\\ \nonumber
d_{3,n}(k) = &\; \frac{1}{4ik^2 + in\omega}\biggl(
- 2\lambda k \sum_{l,m=-\infty}^\infty \bar{a}_{1,l} d_{1,m}(k) d_{1,n+l-m}(k) 
	\\ 
&+ \lambda i \sum_{l, m=-\infty}^\infty  \bar{c}_{1,l} d_{1,m}(k) d_{1,n+l-m}(k)
	\\ \nonumber
& - 2i\lambda \sum_{l,m=-\infty}^\infty \bar{a}_{1,l} a_{1,m} d_{1,n+l-m}(k)
 + 2 k a_{3,n} +i c_{3,n}\biggr), \qquad n \in \Z, \quad k \in \partial D_1^0.
\end{align}
\end{subequations}
This process can be continued indefinitely. Indeed, suppose we have determined $\{c_{M,n}\}_{n=-\infty}^\infty$ and $\{d_{M,n}(k)\}_{n=-\infty}^\infty$ for $N \leq M-1$. Then equation (\ref{cNndef}) with Ê$N = M$ yields
\begin{align*}
c_{M,n} = \bigg(F_{Mn}(k) + 2 ik a_{M,n}\bigg)\bigg|_{k = k_1(n)}, \qquad n \in \Z,
\end{align*}
where the function $F_{Mn}(k)$ is given in terms of known lower order terms:
\begin{align*}
F_{Mn}(k) = &\; \lambda  \sum_{l,m=-\infty}^\infty \bigg\{-2i k \bar{a}_l d_m(k) d_{n+l-m}(k) 
 -   \bar{c}_{l} d_m(k) d_{n+l-m}(k)
	\\
&+ 2 \bar{a}_l a_m d_{n+l-m}(k)\bigg\}_M.
\end{align*}
We can now use equation (\ref{dNndef}) with $N = M$ to determine $d_{M,n}(k)$:
\begin{align*}
d_{M,n}(k) = &\;\frac{1}{4ik^2 + in\omega}\bigl(
F_{Mn}(k) + 2 k a_{M,n} + i c_{M,n} \bigr), \qquad n \in \Z.
\end{align*}
This determines $\{c_{M,n}\}_{n=-\infty}^\infty$ and $\{d_{M,n}(k)\}_{n=-\infty}^\infty$ for $N = M$ and completes the inductive step.

\subsection{Examples}

\begin{example}[Single exponential]\upshape\label{singleexample1}
Suppose 
$$g_0^B(t) = \epsilon e^{i\omega t}, \qquad t \geq 0, \quad \omega > 0.$$
In this case, all coefficients $a_{N,n}$ are zero except for $a_{1,1} = 1$.
Equations (\ref{c1nd1n}) imply that all the coefficients $c_{1,n}$ and $d_{1,n}(k)$ vanish except for
\begin{align*}
  c_{1,1} = - \sqrt{\omega},
\qquad
d_{1,1}(k) = -\frac{i}{2k + i\sqrt{\omega}}.
\end{align*}
Equations (\ref{c2nd2n}) yield $c_{2,n} = d_{2,n}(k) = 0$ for all $n$.
Equations (\ref{c3nd3n}) imply that all the coefficients $c_{3,n}$ and $d_{3,n}(k)$ vanish except for
\begin{align} \nonumber
c_{3,1} = - \frac{\lambda}{2\sqrt{\omega}},
\qquad
d_{3,1}(k) = \frac{\lambda}{2(2ik - \sqrt{\omega})^2 \sqrt{\omega}}.
\end{align}
Continuing in this way, we find that the nonzero coefficients $c_{N,n}$ with $N \leq 8$ are
$$c_{1,1} = - \sqrt{\omega}, \qquad
c_{3,1} = - \frac{\lambda}{2\sqrt{\omega}}, \qquad
c_{5,1} = \frac{1}{8\omega^{3/2}}, \qquad
c_{7,1} = - \frac{\lambda}{16 \omega^{5/2}}.$$
In summary, we have found that
$$g_1^B(t) = -\epsilon \biggl( \sqrt{\omega} + \frac{\epsilon^2 \lambda }{2\sqrt{\omega}} - \frac{\epsilon^4}{8\omega^{3/2}} + \frac{\epsilon^6 \lambda}{16 \omega^{5/2}} + O(\epsilon^8)\biggr)e^{i\omega t}.$$
The summation of this perturbative expansion suggests
\begin{align}\label{g1Bsingleexp}
g_1^B(t) = -\epsilon \sqrt{\omega + \lambda \epsilon^2} e^{i\omega t},
\end{align}
which, upon identifying $\alpha$ and $\epsilon$, is in agreement with (\ref{1.7}a) (note that the inequality $\omega \geq \alpha^2$ is automatically satisfied in the perturbative limit $\alpha \to 0$). 
\end{example}

\begin{example}[Single exponential]\upshape
Suppose 
$$g_{01}^B(t) = \epsilon e^{-i\omega t}, \qquad t \geq 0, \quad \omega > 0.$$
In this case, all coefficients $a_{N,n}$ are zero except for $a_{1,-1} = 1$.
Proceeding as in the previous example, we find that the nonzero coefficients $c_{N,n}$ with $N \leq 8$ are
$$c_{1,-1} = i \sqrt{\omega}, \qquad
c_{3,-1} = - \frac{i \lambda}{\sqrt{\omega}}, \qquad
c_{5,-1} = - \frac{i}{2\omega^{3/2}}, \qquad
c_{7,-1} = - \frac{i \lambda}{2 \omega^{5/2}}.$$
In summary, 
$$g_1^B(t) = i \epsilon \biggl(\sqrt{\omega} - \frac{\epsilon^2 \lambda }{\sqrt{\omega}} - \frac{\epsilon^4}{2\omega^{3/2}} - \frac{\epsilon^6 \lambda}{2 \omega^{5/2}} + O(\epsilon^8)\biggr)e^{-i\omega t}.$$
The summation of this perturbative expansion suggests
$$g_1^B(t) = i \epsilon \sqrt{\omega - 2 \lambda \epsilon^2} e^{-i\omega t},$$
which, upon identifying $\alpha$ and $\epsilon$ and letting $\omega \to -\omega$, is in agreement with (\ref{1.7}b) (note that the inequality $\omega \leq -6\alpha^2$ is automatically satisfied in the perturbative limit $\alpha \to 0$). 
\end{example}

\begin{example}[Sum of exponentials]\upshape\label{sumexample}
We consider the case of
\begin{align}\label{Dirichletcondition}
g_0^B(t) = \epsilon (\alpha e^{i\omega t} + \beta e^{-i\omega t}), \qquad t \geq 0,
\end{align}
where $\alpha, \beta \in \C$ and $\omega > 0$ are constants. In this case, all coefficients $a_{N,n}$ vanish except for 
$$a_{1,1} =  \alpha, \qquad a_{1,-1} = \beta.$$
Equations (\ref{c1nd1n}) imply that all the $c_{1,n}$'s and $d_{1,n}(k)$'s vanish except for
\begin{align*}
c_{1,1} = -\alpha \sqrt{\omega}, \quad 
c_{1,-1} = i \beta \sqrt{\omega}, \quad
d_{1,1}(k) = -\frac{i\alpha}{2k + i\sqrt{\omega}}, \quad
d_{1,-1}(k) = -\frac{i\beta}{2k + \sqrt{\omega}}.
\end{align*}
Equations (\ref{c2nd2n}) yield $c_{2,n} = d_{2,n}(k) = 0$ for all $n$.
In general, $c_{N,n} = 0$ unless both $N$ and $n$ are odd and $N \geq n$.
Continuing in this way, we find that the nonzero coefficients $c_{N,n}$ with $N \leq 8$ are
\begin{subequations}\label{cmnexpressions}
\begin{align}\label{c1expressions}
  c_{1,1} = - \alpha \sqrt{\omega}, \qquad c_{1,-1} = i\beta \sqrt{\omega},
\end{align}  
\begin{align} \nonumber
 & c_{3,3} = -\frac{i (\sqrt{3}+(-2-i)) \alpha ^2 \bar{\beta} \lambda }{2
   \sqrt{\omega }}, &&
  c_{3,1} = -\frac{\alpha  \lambda  (|\alpha|^2 +4 |\beta|^2)}{2 \sqrt{\omega }},
  	\\ \label{c3expressions}
&  c_{3,-1} = -\frac{i \beta  \lambda  (|\beta|^2 +(1-i) |\alpha|^2)}{\sqrt{\omega }}, &&
  c_{3,-3} = \frac{(\sqrt{3}-2-i) \beta ^2 \bar{\alpha} \lambda }{2
   \sqrt{\omega }},
\end{align}
\begin{align} \nonumber
&  c_{5,5} = \frac{((2+i)-(5+2 i) \sqrt{3}-(2+3 i) \sqrt{5}+(3+2 i) \sqrt{15})
   \alpha ^3 \bar{\beta}^2}{16 \omega ^{3/2}}, 
   	\\ \nonumber
&  c_{5,3} = -\frac{i \alpha^2 \bar{\beta} (((3+6 i)+(2+3 i)
   \sqrt{3}) |\alpha|^2+(5+2 i) ((3+2 i)+(2+i)
   \sqrt{3}) |\beta|^2)}{2 ((12+3 i)+(7+2 i)
   \sqrt{3}) \omega ^{3/2}} , 
  	\\\nonumber
&  c_{5,1} = \frac{\alpha  (|\alpha|^4 -4 (\sqrt{3}-6)
   |\alpha|^2  |\beta|^2+2 (9-i   \sqrt{3}) |\beta|^4)}{8 \omega ^{3/2}}, 
   	\\\nonumber
&  c_{5,-1} = \frac{\beta ((\sqrt{3}+(-2+3 i)) |\alpha|^4 - 2 (\sqrt{3}+(2-5 i)) |\alpha|^2  |\beta|^2 -2 i |\beta|^4)}{4 \omega ^{3/2}}, 
  	\\\nonumber
&  c_{5,-3} = -\frac{\beta^2 \bar{\alpha} (((48+9 i)+(29+4 i) \sqrt{3}) |\alpha|^2+8 ((3+5 i)+(2+3 i)   \sqrt{3}) |\beta|^2)}{4 ((21+12 i)+(12+7 i) \sqrt{3}) \omega ^{3/2}} , 
   	\\ \label{c5expressions}
& c_{5,-5} = \frac{((1-2 i)-(2-5 i) \sqrt{3}-(3-2 i) \sqrt{5}+(2-3 i) \sqrt{15})
   \beta ^3 \bar{\alpha}^2}{16 \omega ^{3/2}},
\end{align}
\begin{align} \nonumber 
c_{7,7} = &\; \frac{\alpha ^4 \bar{\beta}^3 \lambda}{32 (56+23
   \sqrt{7}) \omega ^{5/2}}
   \Big\{(-595+147 i)+(525-147 i) \sqrt{3}+(196+49 i) \sqrt{5}
   	\\ \nonumber
&   -(223-30 i) \sqrt{7}-(189-14 i) \sqrt{15}+(192-39 i) \sqrt{21}+(67+19 i) \sqrt{35}
	\\ \nonumber
& -(63+i)  \sqrt{105}\Big\}, 
   	\\ \nonumber
c_{7,5} = &\; \frac{i \alpha ^3 \bar{\beta}^2 \lambda}{8 ((26+97 i)+(15+56 i) \sqrt{3})
   ((25+10 i)+(11+4 i) \sqrt{5}) \omega ^{5/2}}
   	\\ \nonumber
&\times     \Big\{\big((1110+2345 i)+(645+1350 i) \sqrt{3}+(106+1011 i) \sqrt{5}
	\\ \nonumber
& +(63+582 i) \sqrt{15}\big) |\alpha|^2
+(1+i) \big((4542+903 i)+(2623+524 i) \sqrt{3}
	\\ \nonumber
& +(1612+1359 i) \sqrt{5}+(931+786 i) \sqrt{15}\big) |\beta|^2 \Big\}, 
   	\\ \nonumber
 c_{7,3} =  & -\frac{\alpha ^2 \bar{\beta} \lambda}{96 ((627+2340 i)+(362+1351 i)
   \sqrt{3}) \omega ^{5/2}}   \Big\{\big((42588+1476 i)
   	\\ \nonumber
&+(24588+852 i) \sqrt{3}\big) |\alpha|^4 
-(6-6 i) \big((-25328-35261 i)
	\\ \nonumber
& -(14623+20358 i) \sqrt{3}+(3594+1545 i) \sqrt{5}+(2075+892 i) \sqrt{15}\big) |\alpha|^2  |\beta|^2
	\\ \nonumber
& +\big((395885+21770 i)+(228564+12569 i)
   \sqrt{3}-(9405+7020 i) \sqrt{5}
   	\\ \nonumber
&   -(5430+4053 i) \sqrt{15}\big) |\beta|^4 - 96 i \big((795+627 i)+(459+362 i) \sqrt{3}\big)
   |\alpha|^2  |\beta|^2\Big\}, 
  	\\\nonumber
c_{7,1} = & -\frac{\alpha  \lambda }{48 \omega ^{5/2}} 
\Big\{3 |\alpha|^6+6 (4+5 \sqrt{3}) |\alpha|^4  |\beta|^2 +2 \big((162+51 i)
	\\ \nonumber
&-(13+12 i) \sqrt{3}\big) |\alpha|^2 |\beta|^4 
 - 6  |\alpha|^2  |\beta|^2\big((\sqrt{3}-6) |\alpha|^2
   	\\ \nonumber
& +(2+2 i)  (\sqrt{3}+(2+5 i)\big) |\beta|^2)  +16 (15-2 i \sqrt{3}) |\beta|^6\Big\}, 
   	\\\nonumber
c_{7,-1} = &\; \frac{\beta  \lambda}{24 \omega ^{5/2}} \Big\{3 (3-2 i \sqrt{3})|\alpha|^6 +\big((39+42 i)+(28-45 i) \sqrt{3}\big) |\alpha|^4|\beta|^2
	\\ \nonumber
& +((36-15 i)+(23+36 i) \sqrt{3}) |\alpha|^2|\beta|^4 +3
   |\alpha|^2 |\beta|^2 \big((-2+2 i) (\sqrt{3}-6)  |\alpha|^2
   	\\ \nonumber
&   -(\sqrt{3}+(2+5 i)) |\beta|^2 \big)-12 i |\beta|^6\Big\}, 
  	\\\nonumber
c_{7,-3} = &\; \frac{(1+i) \beta^2 \bar{\alpha} \lambda }{96((2340+8733 i)+(1351+5042 i) \sqrt{3}) \omega ^{5/2}} \Big\{3 \big((32088+120795 i)
   	\\ \nonumber
& +(18526+69741 i) \sqrt{3}  +(8106-6393  i) \sqrt{5}+(4680-3691 i) \sqrt{15}\big) |\alpha|^4
   	\\ \nonumber
& +6  i ((168629+32048 i)+(97358+18503 i) \sqrt{3}
 +(8733+11700 i) \sqrt{5}
 	\\ \nonumber
& +(5042+6755 i) \sqrt{15})|\alpha|^2 |\beta|^2 
-(4-4 i) \beta \big(4 ((265+7718 i)
	\\ \nonumber
& +(153+4456 i) \sqrt{3}) |\beta|^2 \bar{\beta}-9 ((892+3329 i)+(515+1922 i)  \sqrt{3}) |\alpha|^2 \bar{\beta}\big)\Big\}, 
   	\\ \nonumber
c_{7,-5} = & -\frac{(1-i) \beta^3 \bar{\alpha}^2 \lambda  }{24((97+26 i)+(56+15 i) \sqrt{3}) ((15+40 i)+(7+18 i) \sqrt{5}) \omega ^{5/2}}
   	\\ \nonumber
&\times  \Big\{3 \big((3820+3375 i)+(2205+1950 i) \sqrt{3}+(1464+1395 i)
   \sqrt{5}
   	\\ \nonumber
&   +(845+806 i) \sqrt{15}\big) |\alpha|^2+(1+i) \big((7755+7962 i)+(4466+4602 i) \sqrt{3}
   	\\ \nonumber
&   +(2742+2586 i) \sqrt{5}+(1578+1495 i) \sqrt{15}\big) |\beta|^2\Big\}, 
   	\\ \nonumber
c_{7,-7} = &\; \frac{\beta ^4 \bar{\alpha}^3 \lambda}{32 (56+23 \sqrt{7}) \omega ^{5/2}}
   \Big\{(-147-595 i)+(147+525 i) \sqrt{3}-(49-196 i) \sqrt{5}
   	\\ \nonumber
&   -(30+223 i) \sqrt{7}-(14+189 i) \sqrt{15}+(39+192 i) \sqrt{21}-(19-67 i) \sqrt{35}
	\\ \label{c7expressions}
& +(1-63  i) \sqrt{105}\Big\} .	
\end{align}
\end{subequations}
\end{example}

\section{Finite-gap solutions and potentially asymptotically admissible pairs}
Finally, we propose an approach for generating pairs $\{g_0^B, g_1^B\}$ of periodic functions which can arise as asymptotic Dirichlet and Neumann values of a solution of NLS. We make the following definition.

\begin{definition}\upshape\label{admissibledef}
A pair of functions $\{g_0^B(t), g_1^B(t)\}$ is {\it asymptotically admissible} if there exists a smooth solution $q(x,t)$ of NLS in the quarter plane $\{x> 0, t > 0\}$, such that 
\begin{enumerate}[(i)]
\item $q(x,t)$ and all its partial derivatives have continuous extensions to $\{x\geq 0, t \geq 0\}$.

\item The initial data decay for large $x$, that is, $q(\cdot,0) \in \mathcal{S}(\R_+)$.
 
\item As $t \to \infty$, the Dirichlet and Neumann boundary values of $q$ asymptote towards $g_0^B(t)$Ê and $g_1^B(t)$ respectively, that is,
\begin{align}\label{qqxasymptote}
q(0, \cdot) - g_0^B \in \mathcal{S}(\R_+), \qquad q_x(0,t) - g_1^B(t) = O(t^{-3/2}), \qquad t\to \infty.
\end{align}
\end{enumerate}
\end{definition}

The spectral functions $\{a(k),b(k),A(k),B(k)\}$ associated with the initial and boundary values $\{q_0(x), g_0(t), g_1(t)\}$ of a solution $q(x,t)$ of (\ref{1.1}) in the quarter plane are not independent, but satisfy an important relation called the {\it global relation}. If the Dirichlet and Neumann values satisfy (\ref{qqxasymptote}) where $\{g_0^B(t), g_1^B(t)\}$ is a pair of smooth periodic functions of period $\tau > 0$, then the global relation requires that the quotient $B(k)/A(k)$ be continuous in a certain region of the complex $k$-plane. In situations where $B(k)/A(k)$ can be explicitly computed, it is possible to use this continuity condition to impose constraints on $\{g_0^B(t), g_1^B(t)\}$. This approach was employed in \cite{BKS2009} to derive the characterization (\ref{1.7}) of asymptotically admissible single exponential boundary values for the focusing NLS equation. In \cite{BKS2009}, the continuity requirement that the global relation imposes on $B(k)/A(k)$ characterizes $\{g_0^B(t), g_1^B(t)\}$ completely.

In general, the nonexplicit nature of the quotient $B(k)/A(k)$ makes it challenging to apply the above approach. However, there exists a large class of pairs $\{g_0^B, g_1^B\}$ for which the quotient $B(k)/A(k)$ is known explicitly. This is the class of finite-gap solutions generated by the Baker-Akhiezer formalism. We propose that potentially asymptotically admissible pairs for the NLS equation can be generated by taking finite-gap solutions for which the data $\{g_0^B, g_1^B\}$ is periodic (not just quasiperiodic) and then enforcing the continuity condition derived from the global relation on $B(k)/A(k)$.

\ack {\it The authors acknowledge support from the EPSRC, UK.}


\end{document}